\documentclass[aps,floatfix,footinbib,a4paper,superscriptaddress,twocolumn,prb,
showpacs]{revtex4}
\usepackage{graphicx}
\usepackage{bm}
\usepackage{amssymb}
\usepackage{color}
\usepackage[normalem]{ulem}

\begin{document}

  \title{Polaron effects on the dc- and ac-tunneling characteristics\\
  of molecular Josephson junctions}

  \author{B. H. Wu}
  \email{bhwu2010@gmail.com}
  \affiliation{Department of Applied Physics, Donghua University, 2999
  North Renmin Road, Shanghai 201620, China}

  \author{J. C. Cao}
  \affiliation{Key Laboratory of Terahertz Solid-State Technology,
  Shanghai Institute of Microsystem and Information
  Technology, Chinese Academy of Sciences, 865 Changning Road,
  Shanghai 200050, China}

  \author{C. Timm}
  \email{carsten.timm@tu-dresden.de}
  \affiliation{Institute of Theoretical Physics, Technische
  Universit\"at Dresden, 01062 Dresden, Germany}

  \date{July 10, 2012}

  \begin{abstract}
    We study the interplay of polaronic effect and superconductivity
    in transport through molecular Josephson junctions. The tunneling rates
    of electrons are dominated by vibronic replicas of the superconducting
    gap, which show up as prominent features in the differential conductance
    for the dc and ac current. For relatively large molecule-lead coupling,
    a features that
    appears when the Josephson frequency matches the vibron frequency can be
    identified with an over-the-gap structure observed by Marchenkov
    {\emph{et al.}}\ [Nat.\ Nanotech.\ \textbf{2}, 481 (2007)].
    However, we are more concerned with
    the weak-coupling limit, where resonant
    tunneling through the molecular level dominates. We find
    that certain features involving both Andreev reflection and
    vibron emission show an unusual shift of the bias voltage $V$ at their
    maximum with the gate voltage $V_g$ as $V\sim (2/3)\,V_g$.
    Moreover, due to the polaronic effect, the ac Josephson current shows
    a phase shift of $\pi$ when the bias $eV$ is
    increased by one vibronic energy quantum $\hbar\omega_v$.
    This distinctive even-odd effect is explained in terms of
    the different sign of the coupling to vibrons of electrons and of
    Andreev-reflected holes.
  \end{abstract}



\maketitle

\section{Introduction}

Electron transport through quantum dots embedded in Josephson junctions is
attracting increasing interest.\cite{NN5703,AP60899} Rich
phenomena\cite{Nature442667,NL72441,NN2481,PRB77195329,
PRB72224502,PRB73214501,PRL101087002} arise due to the competition of
superconductivity and strong interactions. A crucial feature of
molecular Josephson junctions (MJJs) is the interplay of the ac Josephson effect and
molecular vibrations. Signatures in the transport
can be expected when the Josephson frequency matches a vibration frequency.
Since the Josephson frequency can be precisely controlled by the bias voltage,
such signatures could form the basis of a precise molecular spectroscopy.
This idea has recently been
explored by Marchenkov {\emph{et al.}},\cite{NN2481} who have measured the
differential conductance of a Nb dimer in a superconducting break junction. An
over-the-gap structure consisting of a series of peaks was ascribed to the
excitation of vibrational modes by the ac Josephson current,\cite{NN2481} but
no microscopic description was provided. Theoretical treatments of MJJs have
mainly focused on the dc current either for weak
molecule-lead coupling at vanishing bias\cite{PRB72224502} or for weak
electron-vibron coupling\cite{PRB73214501,PRL101087002} so that the microscopic
understanding of spectroscopic signatures in transport through MJJs is still far from
complete. However, such an understanding is the prerequisite for developing the dc and
ac Josephson currents in MJJs into spectroscopic tools.

The purpose of this paper is to investigate the transport properties of
a biased MJJ within a microscopic model. We focus on MJJs with strong
electron-vibron coupling. In this limit, electrons dress with a vibron
cloud, forming polarons. A minimal model of a single-orbital
molecule between two $s$-wave superconducting leads is considered. Electrons in
the molecular orbital are coupled to a vibrational mode of frequency
$\omega_v$. We will see that signatures in transport do not only occur when the
Josephson frequency $\omega_J$ equals the vibration frequency $\omega_v$ but
also when $\hbar\omega_J$, $\hbar\omega_v$, and the superconducting gap in the
leads satisfy certain other simple rational relations.

The remainder of this paper is organized as follows: In Sec.\ \ref{sec.model}
we present our model and discuss the theoretical approach. In Sec.\
\ref{sec.results} we then present and discuss our results for the dc and ac
Josephson effects, followed by a summary in Sec.\ \ref{sec.summary}.

\section{Model and method}
\label{sec.model}

Our model Hamiltonian reads
\begin{eqnarray}
H=\sum_{\alpha= L, R} H_\alpha + H_{\rm{mol}} + H_{T},
\end{eqnarray}
where the first term describes the left (\textit{L}) and right (\textit{R}) BCS
superconducting leads,
\begin{eqnarray}
H_\alpha = \sum_{\mathbf{k}\sigma} \epsilon_{\alpha\mathbf{k}\sigma}\,
  c^\dag_{\alpha\mathbf{k}\sigma} c_{\alpha\mathbf{k}\sigma}
  + \! \sum_{\mathbf{k}\sigma}
  \left( \Delta^*_\alpha\, c_{\alpha\mathbf{k}\uparrow}
  c_{\alpha,-\mathbf{k},\downarrow}\!+\!\mathrm{H.c.} \right)\!,
\end{eqnarray}
with superconducting order parameter $\Delta_\alpha$.
$c_{\alpha\mathbf{k}\sigma}$ ($c^\dag_{\alpha\mathbf{k}\sigma}$) annihilates
(creates) an electron of wave vector $\mathbf{k}$ and spin $\sigma$
in lead $\alpha$.

The molecule with vibration degree of freedom is represented by
\begin{eqnarray}
H_{\rm mol} &=& \sum_\sigma \varepsilon_m
  d^\dag_\sigma d_\sigma + \hbar\omega_{v} a^\dag a
  + \lambda (a^\dag+a) \sum_{\sigma} d^\dag_\sigma d_\sigma, \quad
\label{Hmol}
\end{eqnarray}
where $\varepsilon_m$ is the molecular energy level, $d_\sigma$
($d^\dag_\sigma$) is the annihilation (creation) operator of a spin-$\sigma$
electron in the molecular orbital, $a$ ($a^\dag$) is the vibron annihilation
(creation) operator, and $\lambda$ is the electron-vibron coupling strength.
We neglect the Coulomb interaction in the molecule, which is
justified if the charging energy is small
compared to the coupling to the leads.\cite{EPJB57279} The role of the Coulomb
interaction in MJJs has recently been reviewed in Ref.\ \onlinecite{NN5703}.

The tunneling between the molecule and the leads is described by
\begin{eqnarray}
H_{T} & = & \frac{1}{\sqrt{N}} \sum_{\alpha \mathbf{k}\sigma} t_{\alpha d}\,
  \exp\left[\frac{i}{2}\left(\phi_\alpha +
  \frac{2eV_\alpha}{\hbar}\,t\right)\right]c_{\alpha \mathbf{k}\sigma}^\dag
  d_\sigma \nonumber \\
&& {}+ \mathrm{H.c.},
\end{eqnarray}
where $\phi_\alpha$ is the initial phase of the superconducting order parameter
at time $t=0$, $V_\alpha$ is the voltage in lead $\alpha$, and $t_{\alpha d}$ is
the tunneling matrix element. In the following, we choose $\phi_\alpha=0$,
$V_L=0$, and $V_R=-V$, where $V$ is the voltage drop across the junction. For
symmetric capacitances between the molecule and the leads, as assumed here, the
molecular energy level is then given by $\varepsilon_m=\varepsilon_0-eV/2$.

To go beyond perturbative
approaches,\cite{PRB72224502,PRB73214501,PRL101087002} we employ the unitary
Lang-Firsov transformation\cite{LaF63,Mahan,JPCM19103201} to
diagonalize $H_{\rm mol}$. The transformed Hamiltonian reads
\begin{eqnarray}
\label{TransHamil}
\tilde H=H_\alpha
  + \sum_\sigma \left(\tilde\varepsilon_0-\frac{eV}{2}\right) d^\dag_\sigma
    d_\sigma
  + \hbar\omega_{v}\, a^\dag a +\tilde{H}_{T},
\end{eqnarray}
where the molecular energy level is shifted to $\tilde \varepsilon_0
=\varepsilon_0-{\lambda^2}/{\hbar\omega_{v}}$ by the polaron binding energy.
To simplify notation, we now take $\varepsilon_0$ to denote the shifted level.
In principle, the electron-electron interaction is also renormalized by the
transformation but we neglect this shift together with the bare Coulomb
interaction. We do not expect the on-site interaction to qualitatively change
our results, which are concerned with transport outside of the Coulomb-blockade
regime. $\tilde{H}_{T}$
has the same form as $H_{T}$, except that the tunneling matrix elements are
dressed by the polaronic effect as $\tilde t_{\alpha d}=t_{\alpha d}X$, where
\begin{eqnarray}
X = \exp\left[-\frac{\lambda}{\hbar\omega_{v}}\,(a^\dag-a)\right]
\end{eqnarray}
is the polaron-shift operator.

The transport properties are obtained by the
non\-e\-qui\-li\-bri\-um-Green-func\-tion method.\cite{Mahan,Rammer} In the
Nam\-bu representation, we
introduce the contour-ordered Green function
$G(t,t') = -i\langle T_c \, \psi(t)\psi^\dag(t')\rangle$ with
the contour-ordering directive
$T_c$ and $\psi=(d_{\uparrow}, d^\dag_{\downarrow})^T$. The
particle current through
the lead $\alpha$ is\cite{Rammer,PRB547366,PRB65075315}
\begin{eqnarray}
I_{\alpha}(t) &=& \frac{2e}{\hbar}\,
    {\rm Re}\int dt_1\, \mathrm{Tr}\,\big\{\sigma_z
  \big[ G^<(t,t_1) \Sigma^a_{\alpha}(t_1,t) \nonumber \\
&& {}+ G^r(t,t_1)\Sigma^<_{\alpha}(t_1,t)\big]\big\},
\end{eqnarray}
where the trace is over Nambu space,
$\sigma_z$ is a Pauli matrix, and $\Sigma^a_\alpha$ and
$\Sigma^<_\alpha$ are, respectively, the advanced and lesser
self-energies due to the coupling to lead $\alpha$. The advanced
self-energy is related to the retarded one through
$\Sigma^a(t,t') = [\Sigma^r(t',t)]^\dag$.

The current measured in, say, the left lead consists of the
particle current $I_L(t)$ plus the displacement current due to the formation of
image charges. This contribution vanishes in the stationary
state but must be taken into account in the time-dependent case to ensure gauge
invariance and current conservation. This requires proper partitioning of
the displacement current.\cite{PRL82398} Since we assume symmetric coupling,
the displacement currents are symmetric, in which case the measured current
equals the symmetrized current $I=(I_L-I_R)/2$.\cite{PRL82398,PRB82205112}
Due to the ac Josephson effect, the current can be expanded as
$I(t)=\sum_n I_{n}e^{in\omega_{J} t}$, where the Josephson frequency is
$\omega_{J}=2\,eV/\hbar$ and $I_n=(I_{L,n}-I_{R,n})/2$. The Fourier
components $I_{\alpha,n}$ are
\begin{eqnarray}\label{It}
I_{\alpha,n} & = & \frac{2e}{h}\int d\epsilon\: \mathrm{Re}\,\mathrm{Tr}\sum_m
  \big\{\sigma_z\big[
  G^<_{-n,m}(\epsilon)\, \Sigma^a_{\alpha;m,0}(\epsilon) \nonumber \\
&& {}+ G^r_{-n,m}(\epsilon)\, \Sigma^<_{\alpha;m,0}(\epsilon) \big]\big \},
\end{eqnarray}
where $G_{m,n}(\epsilon) \equiv G_{n-m}(\epsilon+m\omega_J)$ and
\begin{eqnarray}
G_n(\epsilon) \equiv \int_{-\infty}^\infty dt\, \frac{1}{T} \int_0^T dt'\,
  e^{i\epsilon(t-t')}\, e^{-in\omega_Jt'}\, G(t,t')
\end{eqnarray}
and analogously for the self-energies.
The current can be decomposed into dissipative
($I^D_{n}$) and
nondissipative ($I^S_{n}$) contributions,\cite{PRB547366,PRB65075315}
\begin{eqnarray}\label{Jac}
  I(t) = I_{0} + \sum_{n>0}\left( I^D_{n}\cos n\omega_{ J}t
   + I^S_{n}\sin n\omega_{ J}t \right) ,
\end{eqnarray}
where $I^D_{n} = {\rm{Re}}\,(I_{-n}+I_{n})$ and $I^S_{n} = {\rm{Im}}\,
(I_{-n}-I_{n})$.

\begin{figure}
  \includegraphics[width=2.7in]{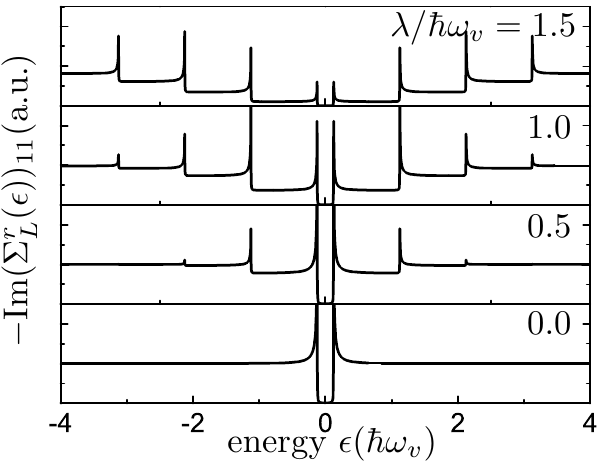}
  \caption{Imaginary part of the first diagonal element of
  the retarded self-energy, $-\mathrm{Im}\,(\Sigma^r_L)_{11}$, which
  represents scattering of electrons between the molecule and lead $L$,
  for various electron-vibron coupling strengths $\lambda$.
  We have taken $\hbar\omega_{v}=8\,\Delta$ and
  $\Gamma=0.2\,\hbar\omega_{v}$.
  With increasing $\lambda$, multiple vibronic replicas of the superconducting
gap
  edges appear.}
  \label{fig.Sigmar}
\end{figure}

Before turning to transport properties, we show that the mo\-le\-cule-lead
coupling is
drastically modified by the polaronic effect.
We employ the standard decoupling approximation, which assumes that
averages of products of two polaron-shift operators
can be taken out of electronic Green functions and evaluated in
equilibrium.\cite{Mahan,PRB71165324,PRB68205323,PRB76033417} This approximation
is valid if $\hbar\omega_v$ or $\lambda$ is large compared to the molecule-lead
coupling. It is known that beyond its range of quantitative
reliability, the approximation predicts resonant features that are too broad and
too low, but that are centered at the same energy as obtained from a more
advanced treatment.\cite{PRB76033417} In contrast to previous studies for normal
leads, we encounter not only the correlation function
$\langle X(t-t')X^\dagger\rangle$ but also $\langle X(t-t')X\rangle$. These
functions are given by\cite{Mahan,JPCM19103201,PRB68205323,PRB76033417}
\begin{eqnarray}
\langle X(t)X^\dag\rangle &=& \langle X^\dag(t) X\rangle
  = \sum_l L_l \, e^{-in\omega_v t},
\label{XX} \\
\langle X(t)X\rangle &=& \langle X^\dag(t) X^\dag\rangle
  = \sum_l (-1)^l\, L_l \, e^{-in\omega_v t} , \quad
\label{XX1}
\end{eqnarray}
where
\begin{eqnarray}
L_l \equiv
  e^{-({\lambda}/{\hbar\omega_{v}})^2(2N_{v}+1)}\,
  \exp\left(\frac{l\hbar\omega_{v}}{2k_BT}\right) \, I_l(\eta),
\end{eqnarray}
with the modified Bessel function $I_l$ of the argument $\eta= 2
({\lambda}/{\hbar\omega_{v}})^2 \sqrt{N_{v}(N_{v}+1)} $. $N_{v}$ is the
average vibron number at temperature $T$ determined by the Bose-Einstein
distribution function. The correlation functions can be evaluated analytically.
While the normal one has the well-known
form\cite{Mahan,JPCM19103201,PRB68205323,PRB76033417}
\begin{eqnarray}
\langle X(t) X^\dag\rangle & = &
  \exp\big({-}g^2\,[(1-e^{-i\omega_{v}t})(N_{v}+1)  \nonumber \\
&& {} + N_{v}(1-e^{i\omega_{v}t})]\big) ,
\label{XXfinal}
\end{eqnarray}
{with $g\equiv \lambda/\hbar\omega_v$}, the anomalous one is
\begin{eqnarray}
\langle X(t) X\rangle & = &
  \exp\big({-}g^2\,[(1+e^{-i\omega_{v}t})(N_{v}+1) \nonumber \\
&& {} + N_{v}(1+e^{i\omega_{v}t})]) .
\label{XXfinal1}
\end{eqnarray}
The two correlation functions differ by a phase shift of half a vibration
period. This phase shift will turn out to be crucial for the polaron effect on
the Josephson current. Its origin can be traced back to the factor of $(-1)^l$
under the sum in Eq.\ (\ref{XX1}). This factor stems from Andreev reflection:
An electron tunneling out of the molecule transmutes into a hole before
it tunnels back in, which couples to the vibron \emph{with the opposite sign},
as seen by inspecting the last term in Eq.\ (\ref{Hmol}).

Under the decoupling approximation, we obtain the retarded and lesser
self-energies due to the coupling to superconducting lead $\alpha$ as
\begin{eqnarray}\label{SER}
\lefteqn{ [\Sigma^r_{{\alpha};mn}]_{ij}(\epsilon)
  = \sum_{l=-\infty}^{\infty} (-1)^{l(i-j)} L_l
   \bigg\{[\tilde \Sigma^r_{{\alpha};mn}]_{ij}(\epsilon^-_l) } \nonumber \\
&& {}+ \frac{1}{2}\, \Big([\tilde \Sigma^<_{{\alpha};mn}]_{ij}(\epsilon^-_l)
   -[\tilde \Sigma^<_{{\alpha};mn}]_{ij}(\epsilon^+_l)\Big) \nonumber \\
&& {}+ \frac{1}{2i}\left(H[[\tilde \Sigma^<_{{\alpha};mn}]_{ij}]
   (\epsilon^-_l)-H[[\tilde \Sigma^<_{{\alpha};mn}]_{ij}]
   (\epsilon^+_l)\right)\! \bigg\}\quad
\end{eqnarray}
and
\begin{eqnarray}\label{SEL}
[\Sigma^<_{\alpha;mn}]_{ij} &=& \sum_{l=-\infty}^{\infty} (-1)^{l(i-j)} L_l
  \, [\tilde\Sigma^<_{\alpha;mn}]_{ij}(\epsilon+l\omega_v),\quad
\end{eqnarray}
respectively, where $\epsilon^\pm_l\equiv\epsilon\pm l\hbar\omega_{v}$, and
$i,j=1,2$ are Nambu indices. Here, $\tilde \Sigma_\alpha$
is the self-energy in the absence of electron-vibron coupling.
Taking the wide-band limit and making use of our choice of vanishing
initial phases $\phi_\alpha$ at time $t=0$, $\tilde \Sigma_\alpha$ is given
by\cite{PRB547366,PRB65075315}
\begin{eqnarray}
\tilde \Sigma^r_{L;mn}(\epsilon) & = &
  -\frac{i}{2}\,\Gamma_L\, \beta_L(\tilde\epsilon_m)\left(\begin{array}{cc}
    1 & -\frac{\Delta_{L}}{\tilde\epsilon_m} \\
    -\frac{\Delta_{L}}{\tilde\epsilon_m} & 1
  \end{array}\right) ,
\end{eqnarray}
\begin{widetext}
\begin{eqnarray}
\tilde \Sigma^r_{R;mn}(\epsilon) & = & -\frac{i}{2}\,\Gamma_R\left(
  \begin{array}{cc}
    \delta_{mn}\, \beta_R(\tilde\epsilon_{m+{1}/{2}}) &
    \delta_{m,n-1}\, \beta_R(\tilde\epsilon_{m+{1}/{2}})\,
    \frac{-\Delta_{R}}{\tilde\epsilon_{m+{1}/{2}}} \\
    \delta_{m,n+1}\, \beta_R(\tilde\epsilon_{m-{1}/{2}})\,
    \frac{-\Delta_{R}}{\tilde\epsilon_{m-{1}/{2}}} &
    \delta_{mn}\, \beta_R(\tilde\epsilon_{m-{1}/{2}})
  \end{array} \right) , \\
\tilde\Sigma_{L;mn}^<(\epsilon) & = &
  i\,\Gamma_L\,\delta_{mn}f(\tilde\epsilon_m)\,
  \tilde\rho_L(\tilde\epsilon_m) \left(
  \begin{array}{cc}
    1 & -\frac{\Delta_{L}}{\tilde\epsilon_m} \\
    -\frac{\Delta_{L}}{\tilde\epsilon_m} & 1
  \end{array}\right) , \\
\tilde\Sigma_{R;mn}^<(\epsilon) & = & i\,\Gamma_R \left(\begin{array}{cc}
    \delta_{mn}\, f(\tilde\epsilon_{m+{1}/{2}})\,
    \tilde\rho_R(\tilde\epsilon_{m+{1}/{2}}) &
    \delta_{m,n-1}\, f(\tilde\epsilon_{m+{1}/{2}})\,
    \tilde\rho_R(\tilde\epsilon_{m+{1}/{2}})\,
    \frac{-\Delta_{R}}{\tilde\epsilon_{m+{1}/{2}}} \\
    \delta_{m,n+1}\, f(\tilde\epsilon_{m-{1}/{2}})\,
    \tilde\rho_R(\tilde\epsilon_{m-{1}/{2}})\,
    \frac{-\Delta_{R}}{\tilde\epsilon_{m-{1}/{2}}} &
    \delta_{mn}\, f(\tilde\epsilon_{m-{1}/{2}})\,
    \tilde\rho_R(\tilde\epsilon_{m-{1}/{2}})
  \end{array} \right) ,
\label{tS.4}
\end{eqnarray}
\end{widetext}
where $\tilde\epsilon_m\equiv \epsilon + m\hbar\omega_J$ and
$f(\epsilon) \equiv (e^{\epsilon/k_BT}+1)^{-1}$ is the
Fermi-Dirac distribution function, where the chemical potential in the left lead
is independent of the bias voltage due to our assumption of $V_L=0$ and has been
absorbed into $\epsilon$. The shifts in the arguments $\tilde\epsilon_{m\pm
1/2}$
of the Fermi functions in Eq.\ (\ref{tS.4}) account for the potential
difference between the leads.
$\Gamma_\alpha \equiv 2\pi |t_{\alpha d}|^2\rho^N_\alpha$ describes the
coupling of the molecule to lead $\alpha$, where $\rho^N_\alpha$
is the density of states in lead $\alpha$ in the normal state. In the
wide-band limit, $\Gamma_\alpha$ is independent of energy. Also, for
symmetric coupling we have $\Gamma_L=\Gamma_R\equiv \Gamma$.
The expressions for $\beta_\alpha$ and $\tilde\rho_\alpha$ read
\begin{eqnarray}
\beta_\alpha(\epsilon) & = & \left\{\begin{array}{ccc}
  \displaystyle\frac{\epsilon}{i\,\sqrt{\Delta_\alpha^2-\epsilon^2}} &
  \mathrm{for} &  \Delta_\alpha>|\epsilon|, \\[3ex]
  \displaystyle\frac{|\epsilon|}{\sqrt{\epsilon^2-\Delta_\alpha^2}} &
  \mathrm{for} & \Delta_\alpha<|\epsilon|,
  \end{array} \right. \\
\tilde\rho_\alpha(\epsilon) & = &
  \theta(|\epsilon|-\Delta_\alpha)\,
  \frac{|\epsilon|}{\sqrt{\epsilon^2-\Delta_\alpha^2}} ,
\end{eqnarray}
respectively. Furthermore,
\begin{eqnarray}
H[F](\omega)\equiv\frac{1}{\pi}\, P\!\!\int d\epsilon\,
  \frac{F(\epsilon)}{\omega-\epsilon}
\end{eqnarray}
is the Hilbert transform of the function $F$, where $P$ denotes the principal
value.  The above results are valid for fast vibron
relaxation so that the averages of bosonic operators can be taken in
equilibrium. Figure \ref{fig.Sigmar} shows
$-{\rm Im}\,(\Sigma^r_L)_{11}$ for various electron-vibron coupling strengths
$\lambda$. Without electron-vibron coupling,
$-{\rm Im}\,(\Sigma^r_L)_{11}$ exhibits the superconducting gap
of the left lead. In the presence of
electron-vibron coupling, $-{\rm Im}\,(\Sigma^r_L)_{11}$ develops
vibronic replicas of the gap edges, separated by integer multiples of
$\hbar\omega_{v}$, which open inelastic
transport channels beyond the usual Andreev
reflection.\cite{PRB547366,PRB65075315} They
enable electrons with energies above the superconducting gap
to undergo Andreev reflection under emission or absorption of vibrons.

\section{Results and discussion}
\label{sec.results}

\subsection{Differential conductance of the dc Josephson current}

\begin{figure}
  \includegraphics[width=3.2in]{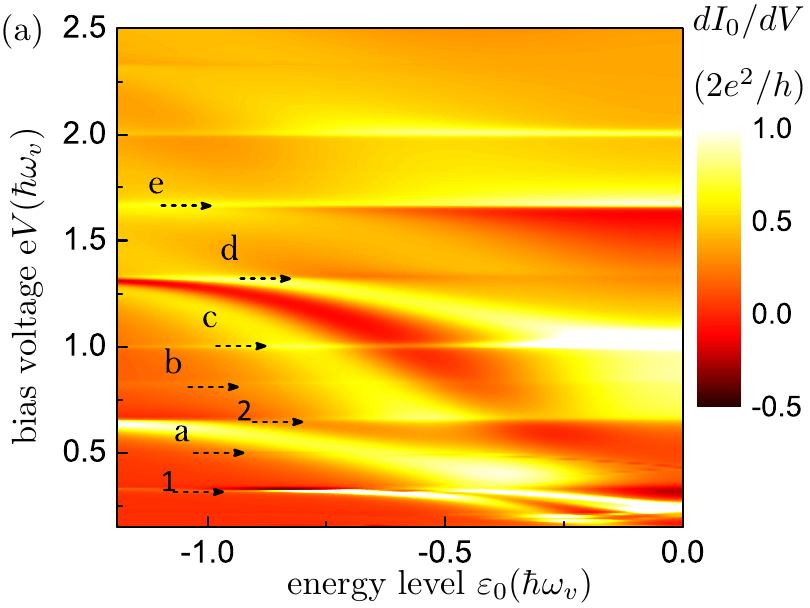}
  \includegraphics[width=3.2in]{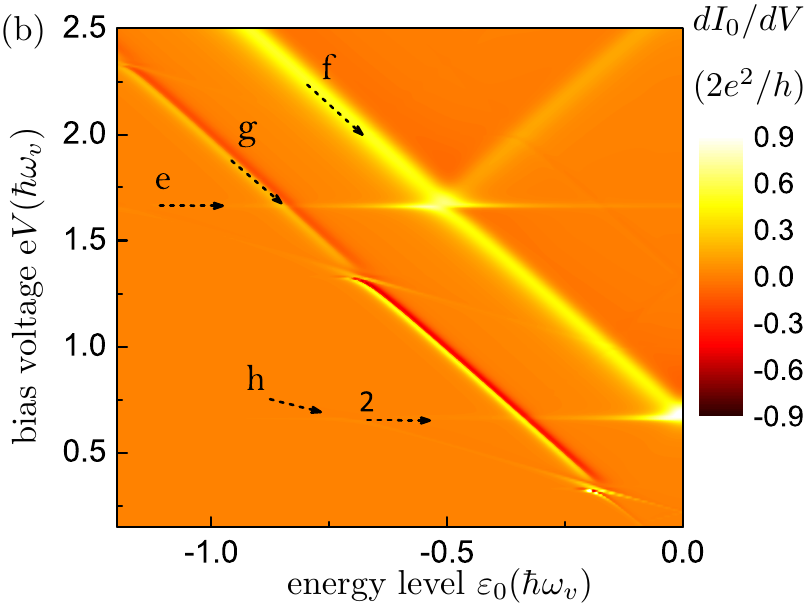}
  \caption{(Color online) Density plot of the dc differential conductance
  $dI_{0}/dV$ vs.\ on-site energy $\varepsilon_0$ and bias
  voltage $V$ for $\Delta=\hbar\omega_v/3$, $\lambda=\hbar\omega_v$,
  and (a) $\Gamma=\hbar\omega_{v}$ and (b) $\Gamma=0.05\,\hbar\omega_{v}$.
  Features labeled ``1'' and ``2'' do not involve vibron emission
  or absorption, whereas those labeled ``a''--``h'' are due to this polaronic
  effect. All features are discussed in the text.}
  \label{fig.dIdV.dc}
\end{figure}

\begin{figure}
  \includegraphics[height=1.0in]{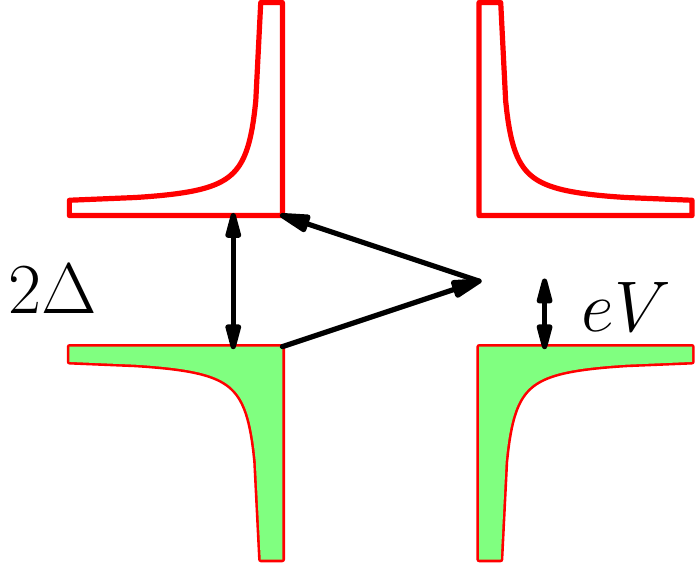}\hspace{2em}
  \includegraphics[height=1.0in]{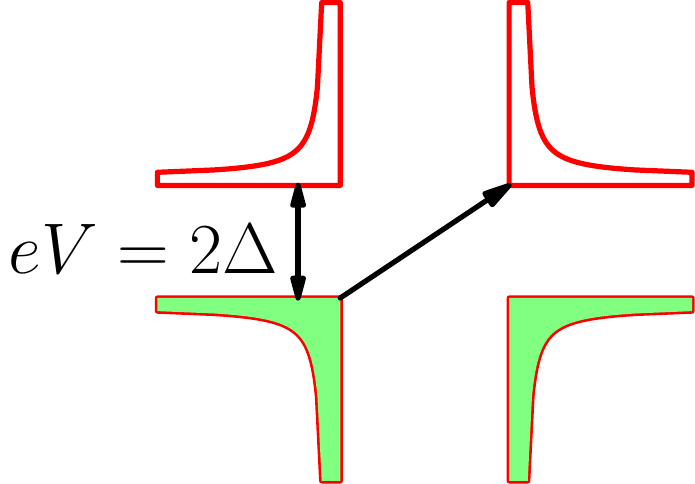}
  \caption{(Color online) Schematic representations of the two most important
  processes in the MJJ that involve neither vibrons nor the molecular level.
  Left panel: single Andreev reflection at the right molecule-lead contact.
  Right panel: direct tunneling from the lower gap edge in the left lead to the
  upper gap edge in the right lead. The filled (empty) shapes represent the
  energy states of the superconducting leads below (above) the Fermi energy.
  Energy scales of the superconducting gap
  $\Delta$ and the bias voltage $eV$ are indicated. The arrows denote possible
  transitions of the electron or hole.}
  \label{fig.simpleproc}
\end{figure}

We now discuss the modification of the Josephson current due
to the polaronic effect. In this study, we assume the two leads to be identical
superconductors and take the amplitude of the order parameters to be
$\Delta \equiv |\Delta_L| = |\Delta_R|$. In Fig.\ \ref{fig.dIdV.dc} we first
present density plots of the dc differential
conductance $dI_{0}/dV$ as a function of the bias voltage $V$ and the on-site
energy $\varepsilon_0$, which in a break-junction setup
could be controlled by a gate voltage. Different
polaronic features are observed depending on the molecule-lead coupling
$\Gamma$. For large
$\Gamma$, dc transport is dominated by coherent tunneling across the molecule
without requiring the energy of the electron to be aligned with the molecular
level. For Fig.\ \ref{fig.dIdV.dc}(a), we have chosen
$\Gamma=\lambda=\hbar\omega_v$. As noted above, in this regime the decoupling
approximation is expected to underestimate the polaronic effect on transport
but to show resonances at the correct voltages.\cite{PRB76033417}
The numerical results thus provide a reasonable qualitative description to the
vibron-assisted tunneling features. In Fig.\ \ref{fig.dIdV.dc}(a), we find
several features at fixed bias voltage, insensitive to $\varepsilon_0$.
The feature labeled by ``1'' at $eV = \hbar\omega_v/3 = \Delta$
is due to single Andreev reflection without vibron emission or absorption. This
process is illustrated by the left panel in Fig.\ \ref{fig.simpleproc}. The
feature labeled by ``2'' at $eV = 2\hbar\omega_v/3 = 2\Delta$ is due to
direct tunneling from the lower gap edge at one side to the upper gap edge at
the other, illustrated by the right panel in Fig.\ \ref{fig.simpleproc}.

\begin{figure}
  \includegraphics[width=0.4\linewidth]{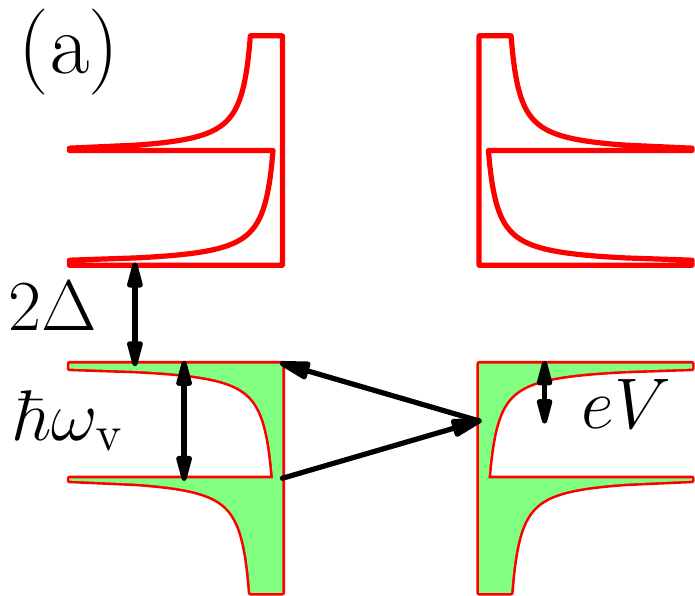}
  \includegraphics[width=0.4\linewidth]{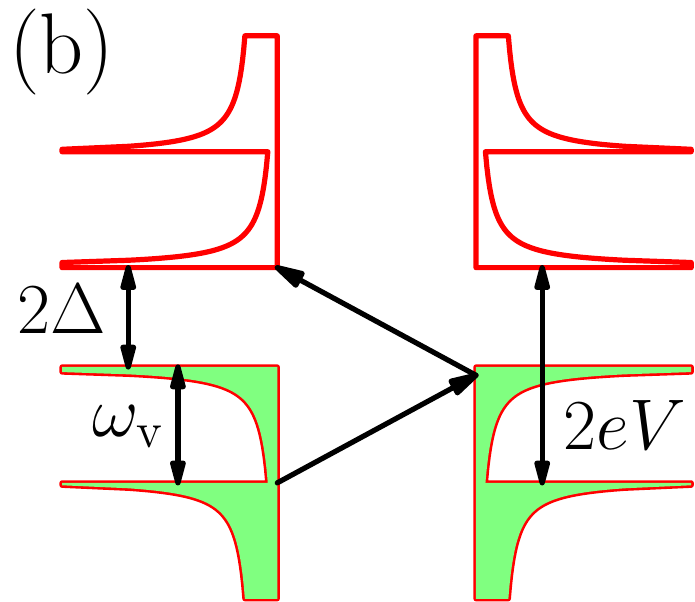}
  \includegraphics[width=0.4\linewidth]{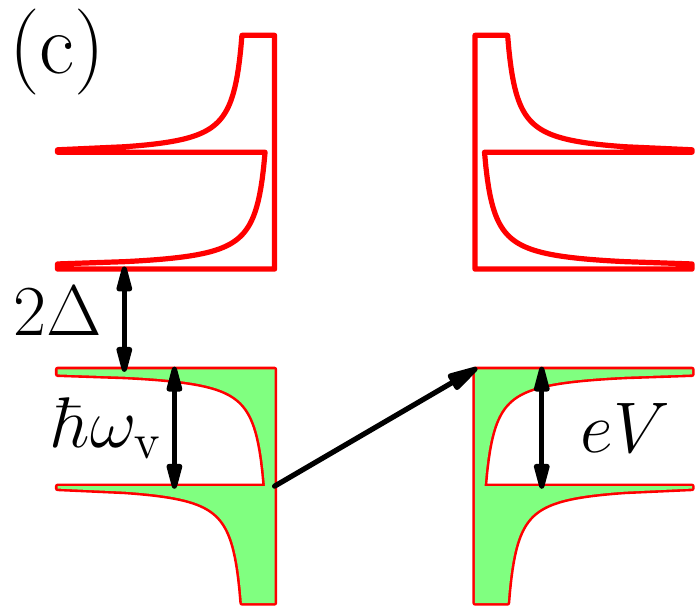}
  \includegraphics[width=0.4\linewidth]{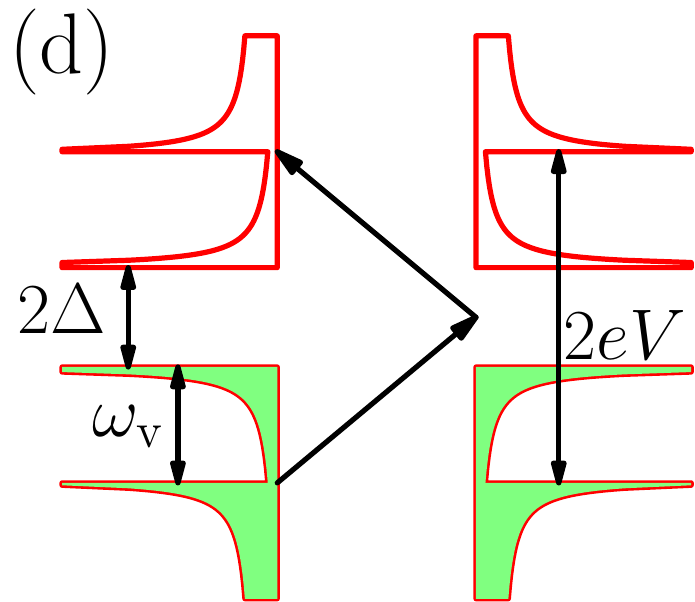}
  \includegraphics[width=0.4\linewidth]{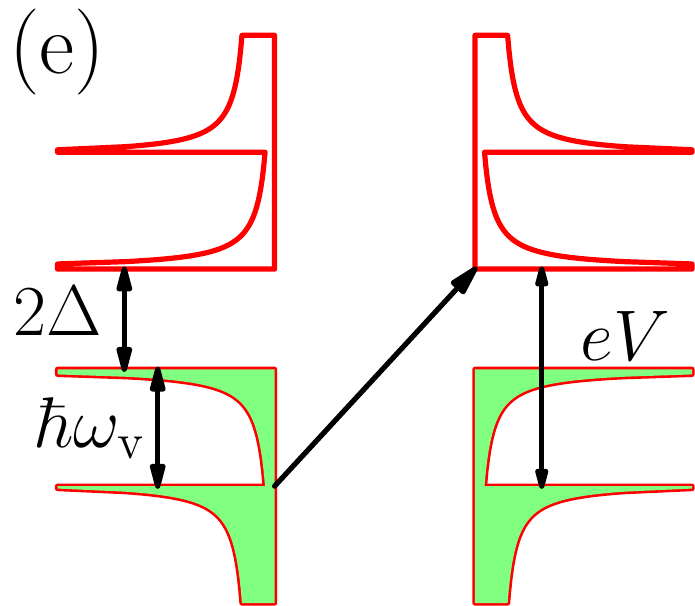}
  \includegraphics[width=0.4\linewidth]{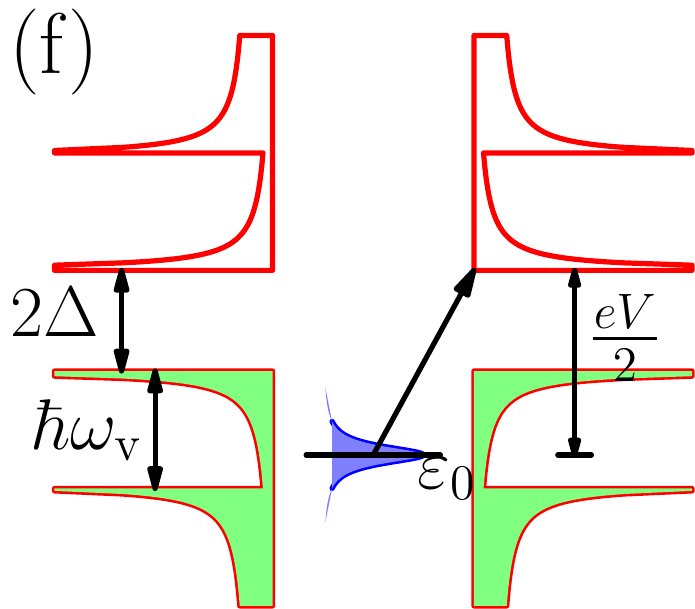}
  \includegraphics[width=0.4\linewidth]{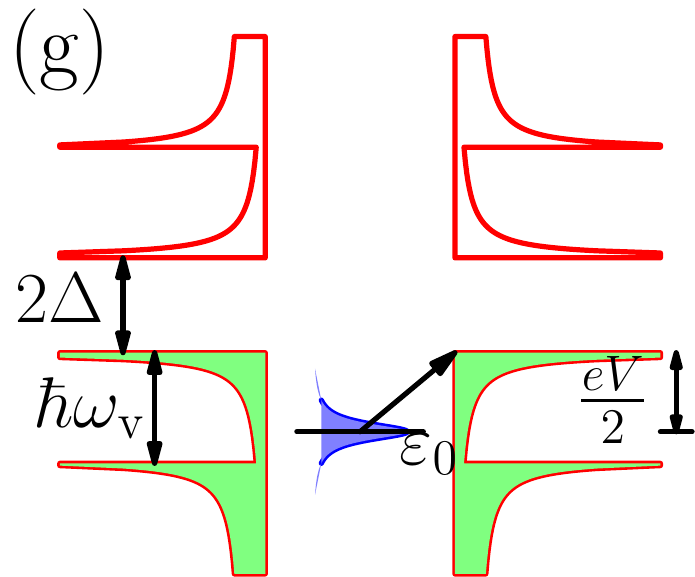}
  \includegraphics[width=0.4\linewidth]{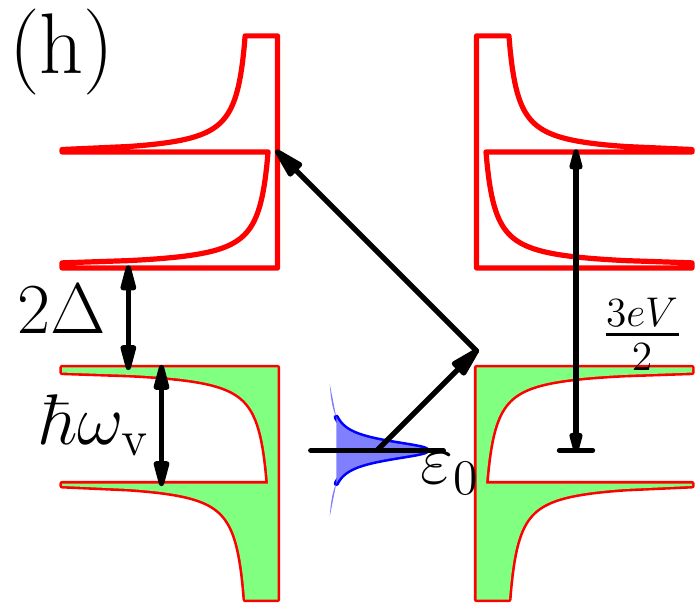}
  \caption{(Color online) Schematic representation of various vibron-assisted
  tunneling and Andreev-reflection processes in the MJJ. The labels are the
  same as in Fig.\ \ref{fig.dIdV.dc}. The DOS of both leads is modified due to
  the polaronic effect. The filled (empty) shapes represent the energy states
  of the superconducting leads below (above) the Fermi energy.
  The molecular level $\epsilon_0$ (blue/gray) is broadened due to the
  coupling to the leads. Energy scales of the superconducting gap
  $\Delta$, the vibron energy $\hbar\omega_v$, and the bias voltage $eV$
  are indicated. The arrows denote possible transitions of the electron or
  hole.}
  \label{fig.procsketch}
\end{figure}

The features labeled by the letters ``a''--``e'' in Fig.\ \ref{fig.dIdV.dc}(a)
involve
vibrons. All these features and their replicas shifted by integer multiples
of $\hbar\omega_v$ can be explained by the onset of
vibron-assisted Andreev reflections
or coherent-tunneling processes sketched in Fig.\ \ref{fig.procsketch},
where the density of states (DOS) has been modified compared to the conventional
picture of Andreev reflection\cite{PRB276739,PRB65075315,PRL91187001}
to account for the polaronic effect. For
instance, the weak feature ``a'' located at $\omega_{v} = \omega_{J}$
arises from the resonance of the vibron and Josephson frequencies and has
been invoked by
Marchenkov {\emph{et al.}}\cite{NN2481} to interpret the observed over-the-gap
structure. The process is sketched in Fig.\ \ref{fig.procsketch}(a), where an
electron from a singular edge of the left lead undergoes one Andreev
reflection at the right lead and arrives at a singular edge of the left lead.
In this process, one vibron is emitted.

On the other hand, for small $\Gamma$, where the decoupling
approximation gives quantitatively reliable results,
the tunneling processes are sensitive to
the position of the molecular energy level. Pronounced features
arise when singular edges of the DOS are aligned with the level $\varepsilon_0$.
In Fig.\ \ref{fig.dIdV.dc}(b), we show $dI_{0}/dV$ vs.\ $V$ and
$\varepsilon_0$ for weak coupling, $\Gamma=0.05\,\hbar\omega_{v}$. As the
non-resonant tunneling is strongly suppressed, the features seen in Fig.\
\ref{fig.dIdV.dc}(a) become much weaker or are even invisible in Fig.\
\ref{fig.dIdV.dc}(b). Instead we see vibron-induced
features with peak bias voltages depending not only on $\omega_v$ and
$\Delta$ but also on
$\varepsilon_0$. Two features, labeled by ``f'' and ``g'', satisfy
$\partial eV/\partial \varepsilon_0 = -2$. They are due to the alignment of
the gap edges with the molecular level
where resonant sequential tunneling through the molecule plays a dominant role.
The corresponding processes are depicted in Figs.\
\ref{fig.procsketch}(f) and (g). Note that the level positions are
renormalized due to the electron-vibron coupling. One can see that feature ``g''
displays a sharp rise in $dI_{0}/dV$ followed by a narrow region of negative
differential conductance. This is due to the onset of resonant tunneling at
the gap edges with singular DOS.\cite{NN5703} Feature ``f''
instead shows a broader structure without negative differential conductance,
since the occupation of the final
state is different, namely nearly empty for ``f'' and nearly full for ``g''.
Moreover, we identify another pronounced feature ``h'' moving with
$\varepsilon_0$, as well as one of its replicas. The underlying picture is shown
in Fig.\ \ref{fig.procsketch}(h). An electron starting from the molecular level
emits a vibron and is Andreev reflected. One could say that the electron
traverses the bias voltage $V$ one and a half times. Accordingly, the feature
has an unusual slope of $\partial eV/\partial \varepsilon_0 = -{2}/{3}$.
Assuming that the molecular level is shifted by a gate voltage $V_g$ as
$\varepsilon_0 = \varepsilon_0' - eV_g$, we predict the distinctive
slope $\partial V/\partial V_g = 2/3$ of this feature in a
bias-voltage/gate-voltage map. Interestingly, the feature is confined
to the voltage range $\hbar\omega_v \le eV \le \hbar \omega_{v}+2\Delta$.
This is due to the fact that for $eV>\hbar\omega_{v}+2\Delta$, electrons prefer
direct tunneling, while for $eV<\hbar\omega_{v}$, the process is blocked due to
the Pauli principle.

\subsection{Differential conductance of the ac Josephson current}

\begin{figure}
  \includegraphics[width=3.2in]{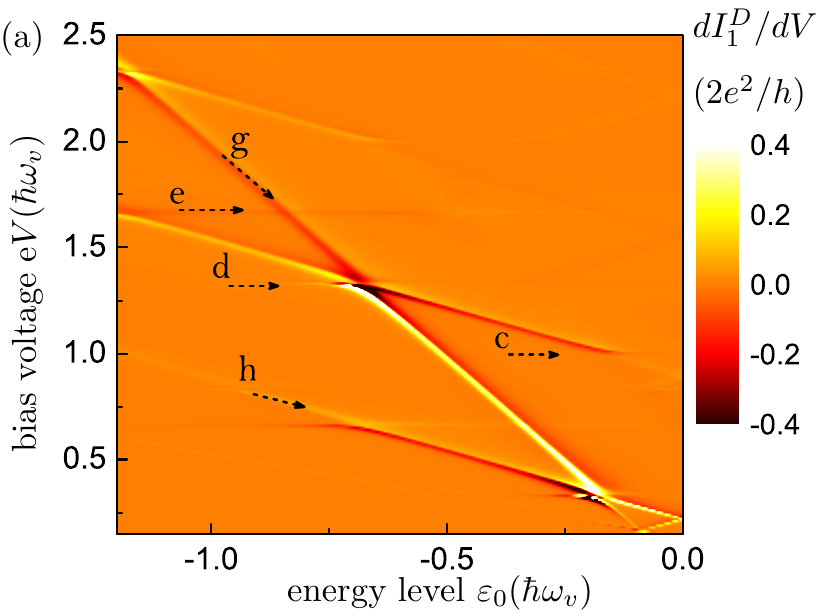}
  \includegraphics[width=3.2in]{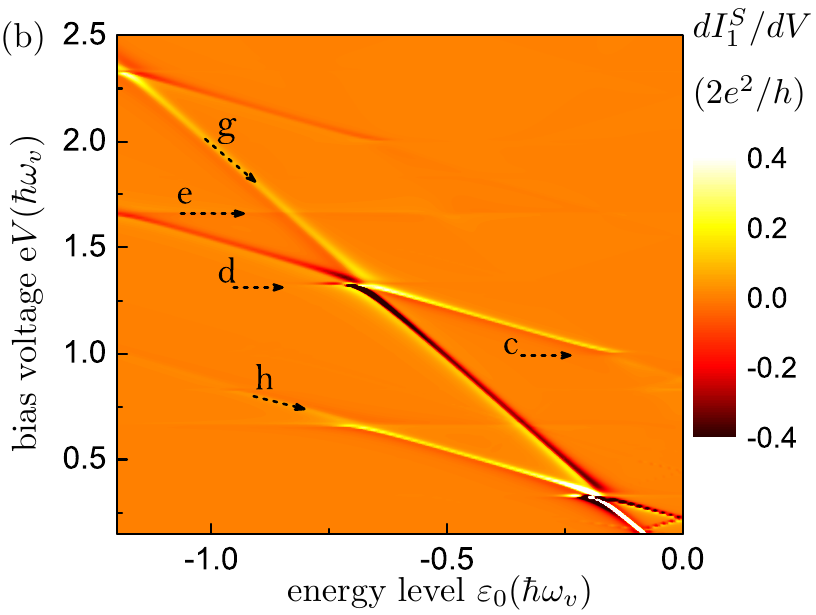}
  \caption{(Color online) Density plot of the ac differential conductance
  of (a) the dissipative Josephson current $I^D_{1}$ and (b) the
  non-dissipative Josephson current $I^S_{1}$ vs.\
  the molecular energy level $\varepsilon_{0}$ and the bias voltage $V$.
  The parameters are identical to those used for Fig.\ \ref{fig.dIdV.dc}(b).}
  \label{fig.dIdV.ac}
\end{figure}

We finally turn to the ac Josephson current. In Fig.\ \ref{fig.dIdV.ac} we plot
the differential conductances of the dissipative and non-dissipative components
as functions of $V$ and $\varepsilon_0$ for small $\Gamma$, where
the decoupling approximation is valid. The features
seen in Fig.\ \ref{fig.dIdV.dc}(b) for the dc current are found again.
However, their appearance is different: Feature ``f'' becomes blurred in the ac
case and feature ``h'' is visible in a much broader voltage range.

More interestingly, we
observe an approximate \emph{antiperiodic} behavior of both the dissipative and
the non-dissipative components of the ac differential conductance as functions
of the bias voltage. The ac current itself exhibits the same antiperiodicity
(not shown). The antiperiod in $eV$ is the vibron energy $\hbar\omega_v$. This
means that the alternating current and the ac differential conductance change
their phase by $\pi$ whenever $eV$ is increased by $\hbar\omega_v$. This
antiperiodicity is a direct
consequence of the polaronic effect: The anomalous (off-diagonal) self-energies
in Eqs.\ (\ref{SER}) and (\ref{SEL}) contain a factor $(-1)^l$.
This factor stems from the corresponding factor in the anomalous correlation
function of the polaron-shift operators, Eq.\ (\ref{XX}). As noted above,
the factor is due to Andreev reflection, since an outgoing electron and
an Andreev-reflected hole couple to the vibron with opposite sign.
For the weak molecule-lead coupling considered here, the current is
dominated by processes involving a single vibron number $l$ determined by $V$.
According to Eqs.\ (\ref{SER}) and (\ref{SEL}), there is then an effective phase
difference across the MJJ of $2eVt/\hbar+l\pi$. The alternating current then
contains a factor of $\sin(2eVt/\hbar+l\pi)$. As a result, the sign of the ac
Josephson current depends on the even-odd parity of the vibron number $l$. The
phase of the ac components can be measured with established
techniques\cite{Barone} or employing the coupling to a charge qubit, as proposed
recently.\cite{PRB84060505} Our results are related to the long-standing
$\cos\varphi$ problem:\cite{Barone,PRB84060505} It was found that the measured
result for the phase of the ac Josephson current does not agree with
theoretical predictions. In the present work we have identified a mechanism by
which this phase could even change periodically as a function of the bias
voltage.

\section{Summary}
\label{sec.summary}

In summary, we have studied the transport properties of MJJs for which the
electronic tunneling rates are modified by polaron formation. Pronounced
features due to its interplay with the superconductivity in the leads
have been identified in the differential conductance of both the dc and ac
Josephson currents. We have explained these features in terms of vibron-assisted
Andreev reflection. The combination of sequential tunneling and Andreev
reflection leads to conductance peaks that show an unusual shift of their peak
bias voltage with the molecular energy level or gate voltage, $V\sim
(2/3)\,V_g$. Furthermore, the opposite sign of the coupling of electrons
and Andreev-reflected holes to vibrons induces periodic phase changes of the ac
components of the Josephson current---their phase changes by $\pi$ when the bias
voltage $eV$ is increased by one vibrational energy quantum $\hbar\omega_v$. We
propose to search for this clear-cut polaronic effect by measuring the ac
Josephson current through molecular junctions.

\acknowledgments

B. H. W. is grateful for the support by the NSFC (Grant No.\ 11074266) and the
Fundamental Research Funds for the Central Universities. J. C. C. is
supported by the 863 Program (Project No.\ 2011AA010205), the NSFC
(Grant Nos.\ 61131006 and 61021064), the Major National Development Project of
Scientific Instrument and Equipment (Grant No.\ 2011YQ150021), the major project
(Project No.\ YYYJ-1123-1), and the Shanghai Municipal Commission of Science and
Technology (Project No.\ 10JC1417000). C. T. acknowledges useful discussions
with P. M. R. Brydon and financial support by the Deutsche
Forschungsgemeinschaft, in part through Research Unit 1154, \textit{Towards
Molecular Spintronics}.

\end{document}